\documentclass[notitlepage, 11pt]{article}
\usepackage{color}
\usepackage{latexsym}
\usepackage{amsmath, amssymb, amsfonts}
\usepackage[toc]{appendix}
\usepackage[makeroom]{cancel}
\numberwithin{equation}{section}

\newtheorem{lemma}{Lemma}

\newtheorem{theorem}{Theorem}
\newtheorem{corollary}{Corollary}

\newtheorem{proposition}{Proposition}

\newtheorem{remark}{Remark}

\newcommand{\beginsec}{
\setcounter{lemma}{0}
\setcounter{theorem}{0}
\setcounter{corollary}{0}
\setcounter{definition}{0}
\setcounter{example}{0}
\setcounter{proposition}{0}
\setcounter{condition}{0}
\setcounter{assumption}{0}
\setcounter{conjecture}{0}
\setcounter{problem}{0}
\setcounter{remark}{0}
}

\newcommand{\noi}{\noindent}

\newcommand{\E}{\mathbb{E}}
\newcommand{\R}{\mathbb{R}}

\newcommand{\la}{\lambda}
\newcommand{\kap}{\kappa}
\newcommand{\sig}{\sigma}

\newcommand{\eps}{\varepsilon}
\newcommand{\ws}{w_s}

\newcommand{\gam}{\gamma}

\newcommand{\FF}{{\mathbb F}}
\newcommand{\PP}{{\mathbb P}}

\newcommand{\calC}{{\cal C}}

\newcommand{\calF}{{\cal F}}

\newcommand{\calL}{{\cal L}}

\newcommand{\skp}{\vspace{\baselineskip}}

\newcommand{\qed}{\hfill $\Box$ \skp}

\newcommand{\Pu}{\rho}
\newcommand{\cAr}{\frac{c-A}{r}}
\newcommand{\cAra}{\left(\frac{c-A}{r}-a\right)}

\newcommand{\cArd}{\left(\frac{c-A}{r}-d\right)}
\newcommand{\Akr}{\frac{A}{\kap-r}}
\newcommand{\dAkr}{\left(d-\Akr\right)}
\newcommand{\lola}{\frac{l}{\la}}
\newcommand{\Pulola}{\left(\Pu-\lola\right)}

\newcommand{\aAkr}{\left(a-\frac{A}{\kap-r}\right)}

\newcommand{\wAkr}{\left(w-\frac{A}{\kap-r}\right)}

\title{Minimizing Lifetime Poverty with a Penalty for Bankruptcy}

\author{Asaf Cohen\thanks{web: https://sites.google.com/site/asafcohentau/, email: shloshim@gmail.com} \hspace{4em} Virginia R.\ Young\thanks{web: http://dept.math.lsa.umich.edu/people/facultyDetail.php?uniqname=vryoung, \hfill \break email: vryoung@umich.edu} \\ \\ \\
Department of Mathematics\\
University of Michigan\\
Ann Arbor, 48109, USA}

\date{\today}

\begin{document}

\maketitle

\begin{abstract}
We provide investment advice for an individual who wishes to minimize her lifetime poverty, with a penalty for bankruptcy or {\it ruin}.  We measure poverty via a non-negative, non-increasing function of (running) wealth.  Thus, the lower wealth falls and the longer wealth stays low, the greater the penalty.  This paper generalizes the problems of minimizing the probability of lifetime ruin and minimizing expected lifetime occupation, with the poverty function serving as a bridge between the two.  To illustrate our model, we compute the optimal investment strategies for a specific poverty function and two consumption functions, and we prove some interesting properties of those investment strategies.

\smallskip


\smallskip


\end{abstract}

\section{Introduction}\label{sec1}
\beginsec

In {\it Scarcity} \cite{MullShafir2013}, Mullainathan and Shafir described how researchers have directly measured the reduction in mental capacity, or {\it bandwidth}, suffered by people who live with scarcity of money, time, or other resources.  This so-called {\it bandwidth tax} is a result of how poverty forces one's mind to focus on dealing with lack of resources.  In this paper, we provide investment advice for an individual who wishes to minimize her lifetime poverty, with a penalty for bankruptcy or {\it ruin}.  We measure poverty via a non-negative, non-increasing function of (running) wealth.  Thus, the lower wealth falls and the longer wealth stays low, the greater the penalty.

In most work concerning poverty, the goal is to measure how well or how poorly income or wealth is spread across a population.\footnote{See \cite{Chakravarty2009} and \cite{Lambert2001} for extensive bibliographies in this line of research.}  In that literature, the focus is on the {\it distribution} of poverty across a group of individuals, not on controlling poverty for a given individual, as we do in this paper.  In other words, our problem is one of micro-economics, not of macro-economics.

This paper is in the spirit of many of those in the collected works of Merton \cite{Merton1992} in that we optimize an objective function for an individual investing in a Black-Scholes market, that is, a market with one riskless asset earning interest at a constant rate and with one risky asset whose price follows geometric Brownian motion.  Whereas the individual in Merton's model seeks to maximize expected utility of consumption and terminal wealth, the individual in our model minimizes expected ``poverty,'' as measured by a non-decreasing function of (running) wealth, not of consumption or of terminal wealth.  In our model, the individual's rate of consumption is given, but she chooses how to invest in order to minimize poverty during her lifetime.

We characterize the optimal investment policy by using the first and second derivatives of the value function (that is, the minimum expected poverty, with a penalty for ruin), which in turn is a solution of an (non-linear) ordinary differential equation.  We prove comparative statistics of the value function for general poverty and consumption functions.  Also, for a specific choice of the poverty function and two types of consumption functions, we compute (semi-)explicit expressions for both the value function and the optimal investment policy.  For these special cases, we use the convex Legendre transform to determine the value function and the corresponding optimal investment policy.

Mathematically, our problem is closely related to those in the goal-seeking literature, such as minimizing the probability of lifetime ruin,\footnote{For an early reference, see Young \cite{Young2004}, and for a more recent reference, see Bayraktar and Zhang \cite{BZ2015}.} maximizing the probability of reaching a bequest goal,\footnote{Bayraktar and Young \cite{BY2015} and Bayraktar et al.\ \cite{BPY2015}.} minimizing expected lifetime occupation (that is, the time that wealth stays below a given level),\footnote{Bayraktar and Young \cite{BY2010}.}, or minimizing the probability of lifetime drawdown or expected lifetime spent in drawdown.\footnote{Chen et al.\ {CLLL2015}, Angoshtari et al.\ \cite{ABY2015a}, and Angoshtari et al.\ \cite{ABY2015b}.}  In fact, this paper generalizes the problems of minimizing the probability of lifetime ruin and minimizing expected lifetime occupation, with the poverty function serving as a bridge between the two, as we discuss in Remark \ref{rem:comp} below.

The remainder of this paper is organized as follows.  In Section \ref{sec2}, we describe the financial model, and we define the problem of minimizing the expectation of a non-negative, non-increasing function of wealth, the so-called {\it poverty function}, with a penalty for ruin.  At the end of that section, we present a verification lemma that we use to solve the optimization problem.  In Section \ref{sec:value}, we prove some properties of the value function for a general poverty function, and in Section \ref{sec3}, we focus on a specific poverty function and two consumption functions.  Section \ref{sec5} concludes the paper.

\section{The model}\label{sec2}
\beginsec

In Section \ref{subsec21}, we present the financial market in which the individual invests, and we define the cost function that the individual wishes to minimize.  Then, in Section \ref{sec2a}, we present a verification lemma that we use to solve the individual's control problem.

\subsection{Background and statement of problem}\label{subsec21}

We study a model of an individual who trades continuously in a Black--Scholes market with no transaction costs. Borrowing and short selling are allowed. The market consists of two assets: a riskless asset and a risky asset. The price of the riskless asset follows the deterministic dynamics
\begin{align}\notag%\label{301}
dX_t = r X_t dt,
\end{align}
in which $r > 0$ is the constant riskless rate of return. The price of the risky asset follows geometric Brownian motion given by
\begin{align}\notag%\label{302}
dS_t &= S_t\left(\mu dt + \sig dB_t\right),
\end{align}
in which $\mu>r$, $\sig>0$, and $(B_t)_{t\ge 0}$ is a standard Brownian motion on a filtered probability space $(\Omega, \calF, \FF = \{ \calF_t \}_{t \ge 0}, \PP)$, in which $\mathcal{F}_t$ is the augmentation of $\sig(B_u\;:\; 0\le u\le t)$.

Let $W_t$ denote the wealth of the individual's investment account at time $t \ge 0$.  Let $\pi_t$ denote the dollar amount invested in the risky asset at time $t \ge 0$.  An investment policy $\{ \pi_t \}_{t \ge 0}$ is {\it admissible} if it is an $\FF$-progressively measurable process satisfying $\int_0^t \pi^2_s \, ds < \infty$ almost surely, for all $t \ge 0$.

We assume that the individual's net consumption rate equals $c(w) - A$, in which $c(w)$ is the rate of consumption when wealth equals $w$, and $A \ge 0$ is the constant rate of income.  In Section \ref{sec3}, we assume that $c(w)$ is a continuous, non-decreasing function of wealth; in Section \ref{sec4}, we consider two specific consumption rates: a constant consumption rate $c(w) = c$, and a proportional consumption rate $c(w) = \kap w$.  Then, the wealth process follows the dynamics
%\begin{align}\notag
%dW_t &= \left(rW_t-c(W_t,t)+A+(\mu-r)\pi(W_t)\right)dt+\sig\pi(W_t)dB_t,\quad t\ge 0,\\\notag
%W_0&= w.
%\end{align}
\begin{align}\notag%\label{303}
\left\{\begin{array}{ll}\notag
               dW_t &= \left[rW_t + (\mu-r)\pi_t  - c(W_t) + A \right]dt + \sig \pi_t dB_t, \quad t \ge 0,
\\ 
                \;\;W_0&= w.
              \end{array}
\right.
\end{align}

Let $\tau_a$ denote the first time that the wealth reaches $a$, which we will refer as the {\it ruin level}, that is, $\tau_a = \inf\{ t \ge 0: W_t \le a \}$. The individual wants to avoid living in poverty and to avoid bankruptcy or ruin during her lifetime.  Let $\tau_d$ be the random time of death of the individual, independent of the Brownian motion driving the risky asset's price process.  We assume that $\tau_d$ is exponentially distributed with hazard rate $\la > 0$, that is, $\PP(\tau_d > t) = e^{- \la t}$.

The individual seeks to minimize the following cost over admissible investment strategies.
\begin{align}\label{304}
J(w; \{ \pi_t \}):&=\E^w \left[ \int_0^{\tau_a\wedge\tau_d} l(W_t) \, dt + \Pu \cdot {\bf 1}_{\{\tau_a\leq \tau_d\}} \right] \\\notag
&= \E^w \left[ \int_0^{\infty} \la e^{-\la t} \left(\int_0^{\tau_a \wedge t} l(W_s) \, ds \right) dt + \Pu \, e^{-\la \tau_a} \right]\\\notag
&= \E^w \left[ \int_0^{\tau_a}e^{-\la t} \, l(W_t) \, dt  + \Pu \, e^{-\la \tau_a} \right] ,
\end{align}
in which $\E^w$ denotes expectation conditional on $W_0 = w$, $l(\cdot)$ is a non-negative, non-increasing function that measures the economic and physical costs of living in poverty, and $\Pu > 0$ is a constant penalty for lifetime ruin.  We call $l(\cdot)$ the {\it poverty function}.  If we were to allow $\Pu < \frac{l(a+)}{\la}$, then the individual might find it advantageous to commit financial suicide by allowing her wealth to fall to the ruin level instead of continuing to live in poverty.  Therefore, to prevent financial suicide, we assume that $\Pu \ge \frac{l(a+)}{\la}$ throughout this paper.  One can interpret the difference $\Pu - \frac{l(a+)}{\la}$ as the {\it net} penalty for ruin, that is, net of the penalty for ruining instead of remaining in poverty (near $a$) for the rest of one's life.  Furthermore, we assume that $l(a+) > 0$; otherwise, $l(\cdot) \equiv 0$, and our problem would be equivalent to minimizing the probability of lifetime ruin.

The function $V$ defined by 
\begin{align}\label{305b}
V(w):= \inf_{\{\pi_t \}} J(w; \{ \pi_t \}) 
\end{align}
is the {\it value function}, in which we minimize over admissible investment strategies.

\begin{remark}\label{rem:comp}
In \cite{BY2010}, Bayraktar and Young study a special case of the problem in this paper, the so-called {\rm lifetime occupation} problem.  They minimize the expected time that wealth spends below $0$, subject to the ``game'' ending if wealth falls below some very low level, $-L$ in \cite{BY2010}.  Thus, if one sets $\Pu = 1/\la$, $l(w) = {\bf 1}_{\{ w < 0 \}}$, and $a = -L$, then $V$  in \eqref{305b} plus the pre-existing time spent below $0$ equals the minimum lifetime occupation as defined in \cite{BY2010}.  Note that by setting $l(w) = {\bf 1}_{\{ w \le 0\}}$, we measure the running time that wealth spends below $0$, and by setting $\Pu = 1/\la$, we assume that once wealth reaches $a = -L$, then the individual spends the remainder of her life in poverty, with expected time $1/\la$.

One can define {\rm occupation} more generally to mean the time spent in a given interval, say, $[a, d]$, for ruin level $a$.  If one considers poverty functions with support lying in $[a, d]$, then the problem in this paper connects two seemingly unrelated problems: minimizing the probability of lifetime ruin with ruin level $a$ as defined in Young \cite{Young2004} and minimizing expected lifetime occupation of $[a, d]$.  Indeed, at one extreme, if we set $\Pu = 1/\la$ and $l(\cdot) \equiv 0$ on $[a, d]$, then $V$ \eqref{305b} equals $1/\la$ times the minimum probability of lifetime ruin.  The corresponding optimal investment strategy is not changed by this scaling.  At the other extreme, if given $\Pu = 1/\la$, we set $l(\cdot)$ as large as possible on $[a, d]$, specifically, $l(w) = {\bf 1}_{\{ a \le w \le d \}}$, then $V$ \eqref{305b} equals the minimum lifetime occupation of the interval $[a, d]$.  Again, we assume that if wealth reaches $a$, then the individual spends the remainder of her life in poverty, with expected time $1/\la$.
\end{remark}

\begin{remark}
Throughout this paper, we assume that the rate of consumption $c(\cdot)$ is a continuous, non-negative, non-decreasing function of wealth on $(a, \infty)$ and that the poverty function $l(\cdot)$ is a non-negative, non-increasing function of wealth on $(a, \infty)$ with finitely many points of discontinuity, such that there exists a unique $\ws \in (a, \infty]$ for which
\begin{equation*}
rw < c(w) - A, \quad \hbox{ for all } w < \ws,
\end{equation*}
\begin{equation*}
rw > c(w) - A, \quad \hbox{ for all } w > \ws,
\end{equation*}
and
\begin{equation*}
l(w) = 0, \quad \hbox{ for all } w > \ws.
\end{equation*}
We allow $\ws = \infty$, which is the case, for example, if $c(w) = \kap w$ with $\kap > r$.  If $W_0 = w \ge \ws$, then we can set $\pi_t = 0$ for all $t \ge 0$, which implies
\begin{equation*}
dW_t = (rW_t - c(W_t) + A) \, dt \ge 0.
\end{equation*}
Under this investment strategy, the wealth is non-decreasing, so lifetime ruin cannot occur nor will the individual incur penalty for being in poverty.  For this reason, we call $\ws$ the {\rm safe level}.
\end{remark}

\subsection{Verification lemma}\label{sec2a}

In this section, we provide a verification lemma that characterizes the value function as a unique solution to a boundary-value problem. We do not prove the theorem because its proof is similar to others in the literature; see, for example, \cite{Bayraktar2007b}.  For every $\pi \in \R$, define the following differential operator $\calL^\pi$ by
\begin{align}\notag%\label{306}
\mathcal{L}^\pi f := (rw + (\mu-r) \pi - c(w) + A) f_w + \frac{1}{2}\sig^2 \pi^2 f_{ww} - \la f,
\end{align}
in which $f$ is a twice-differentiable function.
 
\begin{lemma}\label{lem:verf}
Let $K = K(w)$ be a $\calC^2$ function that is decreasing and convex on $[a, \ws]$ $($except at points of discontinuity of the poverty function $l(\cdot)$, where it will be $\calC^1$ and have left- and right-derivatives$)$. Suppose $K$ satisfies the following boundary-value problem.\footnote{When $\ws = \infty$, the boundary condition $K(\infty)=0$ should be understood as $\lim_{w\to\infty} K(w)=0$.}
\begin{align}\label{307}
\left\{\begin{array}{ll}
               \displaystyle \inf_{\pi}\{\mathcal{L}^\pi K(w)+l(w)\}=0, \quad a < w < \ws,
\\ \\
              K(a) = \Pu, \quad K(\ws)=0.
              \end{array}
\right.
\end{align}
Then, the value function $V$ defined by \eqref{305b} equals $K$ on $[a, \ws]$, and the optimal amount invested in the risky asset is given in a feedback form by
\begin{align}\notag%\label{308}
\pi^*_t = -\frac{\mu-r}{\sig^2}\cdot\frac{K_w(W^*_t)}{K_{ww}(W^*_t)}
\end{align}
for all $t\in[0,\tau_a\wedge \tau_d]$, in which $W^*_t$ is optimally controlled wealth at time $t$.
\end{lemma}

\section{Properties of the value function}\label{sec:value}
\beginsec

In this section, we assume that, for a given rate of consumption $c(\cdot)$ and poverty function $l(\cdot)$, the boundary-value problem (BVP)\eqref{307} has a decreasing, convex solution ($\calC^2$ except at discontinuities of the poverty function, where it is $\calC^1$).  When we look at specific functional forms for $c(\cdot)$ and $l(\cdot)$ in Section \ref{sec3}, we construct the solution of the BVP, but in this section, we take the solution as given.  Moreover, without ambiguity, we write $V$ for that solution because the verification lemma, Lemma \ref{lem:verf}, ensures us that an appropriate solution of \eqref{307} equals the value function.

We prove properties of the value function in this section.  To that end, define the differential operator $F$ by
\begin{equation}\label{eq:F}
F(w, K, K_w, K_{ww}) = \la K - \inf_{\pi} \left[ (rw + (\mu - r)\pi - c(w) + A)K_w + \frac{1}{2} \, \sig^2 \pi^2 K_{ww} \right] - l(w),
\end{equation}
and use $F$ to rewrite the BVP \eqref{307} as follows.
\begin{align*}
\left\{\begin{array}{ll}
F(w, K, K_w, K_{ww}) = 0, \\
K(a) = \Pu, \quad K(\ws) = 0.
\end{array}
\right.
\end{align*}
Note that $F$ is increasing with $K$ and decreasing with $K_{ww}$.  These monotonicity properties allow us to prove the following comparison lemma.

\begin{lemma}\label{lem:compF}
Let $u, v \in \calC^2(a, \ws)$, except at the points of discontinuity of the poverty function $l(\cdot)$, at which $u$ and $v$ are $\calC^1$ with left- and right-derivatives.  Suppose
\begin{equation}\label{ineq:compF}
F(w, u, u_w, u_{ww}) \le F(w, v, v_w, v_{ww}),
\end{equation}
with $F(w, u, u_w, u_{ww}) < \infty$, for all $w \in (a, \ws)$.\footnote{At a point $d$ of discontinuity of $l(\cdot)$, we require \eqref{ineq:compF} to hold at $w = d-$ and $w = d+$.}  If $u(a) \le v(a)$ and $u(\ws) \le v(\ws)$, then $u(w) \le v(w)$ for all $w \in (a, \ws)$.
\end{lemma}

\noi \textbf{Proof.}
First, if the maximum of $u - v$ occurs on the boundary of $(a, \ws)$, but not in the interior, then $u \le v$ in the interior because $u \le v$ on the boundary by assumption.  Second, if $u - v$ attains a non-positive maximum in the interior of $(a, \ws)$, then we also have $u \le v$ in the interior.

Third, suppose $u - v$ attains a strictly positive maximum at $w_0 \in (a, \ws)$.  If $w_0$ is a point of continuity of the poverty function, then $u_w(w_0) = v_w(w_0)$ and $u_{ww}(w_0) \le v_{ww}(w_0)$.  Because $F(w_0, u(w_0), u_w(w_0), u_{ww}(w_0)) < \infty$, either $u_{ww}(w_0) > 0$ or $u_w(w_0) = u_{ww}(w_0) = 0$.  In the former case, we have
\begin{align*}
0 &\le F(w_0, v(w_0), v_w(w_0), v_{ww}(w_0)) - F(w_0, u(w_0), u_w(w_0), u_{ww}(w_0)) \\
&= - \la (u(w_0) - v(w_0)) - m \, \frac{u_w^2(w_0) (v_{ww}(w_0) - u_{ww}(w_0))}{u_{ww}(w_0) v_{ww}(w_0)} < 0,
\end{align*}
a contradiction.  In the latter case, we have $v_w(w_0) = 0$ and $v_{ww}(w_0) \ge 0$; thus,
\begin{align*}
0 &\le F(w_0, v(w_0), v_w(w_0), v_{ww}(w_0)) - F(w_0, u(w_0), u_w(w_0), u_{ww}(w_0)) \\
&= - \la (u(w_0) - v(w_0)) < 0,
\end{align*}
a contradiction.  If $w_0$ is a point of discontinuity of the poverty function, then the above argument applies with $w_0$ replaced by $w_0-$ or $w_0+$.  \qed

In the next proposition, we apply Lemma \ref{lem:compF} to show how the value function $V$ in \eqref{305b} changes with respect to the various inputs of the model.  Many of the properties of $V$ follow directly from its definition; however, we find it instructive to show the interested reader how to prove these properties using the comparison lemma.  In the proposition, when we write ``increases'' or ``decreases,'' we mean in the weak  or non-strict sense.

\begin{proposition}
\item{$($i$)$} $V$ increases with the penalty for ruin $\Pu$.
\item{$($ii$)$} $V$ increases with the poverty function $l$.  Specifically, if $l_1(w) \le l_2(w)$ for all $w \in (a, \ws)$ $($with $l_1(w) = 0 = l_2(w)$ for all $w > \ws)$, then $V(\cdot \,; l = l_1) \le V(\cdot \,; l = l_2)$.
\item{$($iii$)$} $V$ increases with the consumption function $c$.  Specifically, if $c_1(w) \le c_2(w)$ for all $w \in \left(a, \ws^1 \right)$, in which $\ws^1$ is the safe level for $c_1$, then $V(\cdot \, ; c = c_1) \le V(\cdot \,; c = c_2)$ on $\left(a, \ws^1 \right)$.
\item{$($iv$)$} $V$ decreases with the hazard rate $\la$.
\item{$($v$)$} $V$ decreases with the drift of the risky asset $\mu$.
\item{$($vi$)$} $V$ increases with the volatility of the risky asset $\sig$.
\end{proposition}

\noi \textbf{Proof.}
To prove (i), let $0 < \Pu_1 \le \Pu_2$, and define $V_i(\cdot) := V(\cdot \, ; \Pu = \Pu_i)$ and $F_i(\cdot) := F(\cdot \, ; \Pu = \Pu_i)$ for $i = 1, 2$.  We have
\begin{equation*}
F_1(w, V_2, (V_2)_w, (V_2)_{ww}) = 0 = F_1(w, V_1, (V_1)_w, (V_1)_{ww}),
\end{equation*}
$V_1(a) = \Pu_1 \le \Pu_2 = V_2(a)$, and $V_1(\ws) = 0 = V_2(\ws)$.  Thus, by Lemma \ref{lem:compF}, it follows that $V_1 \le V_2$.

To prove (ii), let $l_1(w) \le l_2(w)$ for all $w \in (a, \ws)$, and define $V_i(\cdot) := V(\cdot \,; l = l_i)$ and $F_i(\cdot) := F(\cdot \,; l = l_i)$ for $i = 1, 2$.  We have
\begin{equation*}
F_1(w, V_2, (V_2)_w, (V_2)_{ww}) = l_2(w) - l_1(w) \ge 0 = F_1(w, V_1, (V_1)_w, (V_1)_{ww}),
\end{equation*}
$V_1(a) = \Pu = V_2(a)$, and $V_1(\ws) = 0 = V_2(\ws)$.  Thus, by Lemma \ref{lem:compF}, it follows that $V_1 \le V_2$.

The proofs of Properties (iii) - (vi) are similar, so we leave them to the reader.  \qed

\begin{remark}
Properties $(i)$ - $(iii)$ make intuitive sense and follow easily from the definition of $V$.  Property $(iv)$ also follows from the definition of $V$; specifically, if $\la$ increases, then the discount function $e^{-\la t}$ in \eqref{304} is smaller for all $t \ge 0$.  From an intuitive point of view, if $\la$ increases, then the individual is more likely to die and thereby has less opportunity to spend time in poverty or to ruin.  We expect Properties $(v)$ and $(vi)$ to hold because increasing $\mu$ or decreasing $\sig$ means that the risky asset is a more effective investment for avoiding poverty and ruin.
\end{remark}

In defining the value function $V$, if we were to set $l \equiv 0$, then the value function would equal the minimum probability of lifetime ruin (times $\Pu$), as defined in \cite{Young2004}.  Thus, as $l \to 0$, we expect $V$ to approach the minimum probability of lifetime ruin, which is indeed the case, as we prove with the help of the following lemma.

\begin{lemma}\label{lem:compFeps}
Fix $\eps>0$. Let $u, v \in \calC^2(a, \ws)$, except at the points of discontinuity of the poverty function $l(\cdot)$, at which $u$ and $v$ are $\calC^1$ with left- and right-derivatives.  Suppose
\begin{equation}\label{ineq:compFeps}
\sup_{w\in[a,w_s]}|F(w, u, u_w, u_{ww}) - F(w, v, v_w, v_{ww})| < \eps,
\end{equation}
with $F(w, u, u_w, u_{ww}) < \infty$, for all $w \in (a, \ws)$.\footnote{At a point $d$ of discontinuity of $l(\cdot)$, we require \eqref{ineq:compFeps} to hold at $w = d-$ and $w = d+$.}  If $u(a) = v(a)$ and $u(\ws) = v(\ws)$, then $|u(w) - v(w)| < \eps/\la$ for all $w \in (a, \ws)$.
\end{lemma}

\noi \textbf{Proof.}
If $u=u$ on $[a, \ws]$, then the proof is done.  We now treat the case for which $u \ne v$ on $[a, \ws]$.  Without loss of generality we may assume that $u - v$ attains a strictly positive maximum at $w_0 \in (a, \ws)$.  If $w_0$ is a point of continuity of the poverty function, then $u_w(w_0) = v_w(w_0)$ and $u_{ww}(w_0) \le v_{ww}(w_0)$.  Because $F(w_0, u(w_0), u_w(w_0), u_{ww}(w_0)) < \infty$, either $u_{ww}(w_0) > 0$ or $u_w(w_0) = u_{ww}(w_0) = 0$.  In the former case, we have
\begin{align*}
-\eps &\le F(w_0, v(w_0), v_w(w_0), v_{ww}(w_0)) - F(w_0, u(w_0), u_w(w_0), u_{ww}(w_0)) \\
&= - \la (u(w_0) - v(w_0)) - m \, \frac{u_w^2(w_0) (v_{ww}(w_0) - u_{ww}(w_0))}{u_{ww}(w_0) v_{ww}(w_0)} < - \la (u(w_0) - v(w_0));
\end{align*}
therefore, $u(w_0) - v(w_0) < \eps/\la$. In the latter case, we have $v_w(w_0) = 0$ and $v_{ww}(w_0) \ge 0$; thus,
\begin{align*}
-\eps &\le F(w_0, v(w_0), v_w(w_0), v_{ww}(w_0)) - F(w_0, u(w_0), u_w(w_0), u_{ww}(w_0)) \\
&= - \la (u(w_0) - v(w_0)),
\end{align*}
and again $u(w_0) - v(w_0) < \eps/\la$.   If $w_0$ is a point of discontinuity of the poverty function, then the above argument applies with $w_0$ replaced by $w_0-$ or $w_0+$.  \qed

In the next proposition, we apply Lemma \ref{lem:compFeps} to show that the value function $V$ in \eqref{305b} is continuous with respect to the poverty function.

\begin{proposition}\label{prop:Vcont}
 $V$ is continuous with respect to the poverty function $l$.  Specifically, if $|l_1(w) - l_2(w)|<\eps$ for all $w \in (a, \ws)$ $($with $l_1(w) = 0 = l_2(w)$ for all $w > \ws)$, then
 \begin{equation*}
 \left|V(\cdot \,; l = l_1) - V(\cdot \,; l = l_2) \right| < \eps/\la,
 \end{equation*}
 uniformly on $[a, \ws]$.
\end{proposition}

\noi \textbf{Proof.}
Suppose $|l_1(w) - l_2(w)| < \eps$ for all $w \in (a, \ws)$, and define $V_i(\cdot) := V(\cdot \,; l = l_i)$ and $F_i(\cdot) := F(\cdot \,; l = l_i)$ for $i = 1, 2$.  We have
\begin{equation*}
| F_1(w, V_1, (V_1)_w, (V_1)_{ww}) - F_1(w, V_2, (V_2)_w, (V_2)_{ww})| = | 0 - (l_1(w) - l_2 (w)) | < \eps.
\end{equation*}
Moreover, $V_1(a) = \Pu = V_2(a)$, and $V_1(\ws) = 0 = V_2(\ws)$.  Thus, by Lemma \ref{lem:compFeps}, it follows that $|V_1  -V_2| < \eps/\la$ on $[a, \ws]$.  \qed

The next corollary follows immediately from Proposition \ref{prop:Vcont}.

 \begin{corollary}
 Let $V^0$ denote the minimum the probability of lifetime ruin times $\Pu$.  Then,
 \begin{equation*}
 \lim_{\| l \|_\infty \to 0} V(w) = V^0(w),
 \end{equation*}
 for all $w \in [a, \ws]$, in which $\| l \|_\infty = \sup_{w\in[a,w_s]} | l(s) |$.
 \end{corollary}

\section{Piecewise constant poverty function}\label{sec3}
\beginsec

In this section, we focus on a specific poverty function and two consumption functions.  The poverty function is given by 
\begin{align}\label{304b}
l(w)=\left\{\begin{array}{ll}
               \displaystyle l,
               &\ a\leq w \le d,
\\
                0, & \  w > d,
              \end{array}
\right.
\end{align}
in which $0 < l \le \Pu \la$ is the running cost for living in poverty, and $d > a$ is the {\it poverty level}.  One can think of $l(\cdot)$ given in \eqref{304b} as the building block for non-negative, non-increasing functions. Indeed, one can write any such function as the pointwise limit of an increasing sequence of functions of the form
\begin{equation*}
l_n(w) = b_0 + \sum_{i=1}^{m_n} b_i \, {\bf 1}_{\{a \le w \le d_i \}};
\end{equation*}
see, for example, Royden \cite{Royden1968}.

In Section \ref{sec30}, we assume that rate of consumption is constant; in Section \ref{sec4}, we assume that the rate of consumption is proportional to wealth.  Our work in those two sections easily extends to the case for which $l(\cdot)$ is given by \eqref{304b} and the rate of consumption is piecewise linear in wealth.  Furthermore, we expect the qualitative results in Sections \ref{sec3c} and \ref{sec4c} concerning the optimal investment strategies to extend to the more general case for which $l(\cdot)$ is any non-negative, non-increasing function, as we discuss in Section \ref{sec5}.

\subsection{Constant rate of consumption}\label{sec30}

In this section, we solve the problem when the rate consumption is constant, that is, $c(w) \equiv c$.  We assume that $c > A$; otherwise, the minimization problem is trivial.  The safe level $\ws$ equals $\cAr$.  In this case, we assume that $d < \cAr$, that is, the poverty level is less than the safe level.  We first analyze an auxiliary free-boundary problem (FBP).  Then, in Section \ref{sec3b}, via the Legendre transform, we connect the FBP with the problem of minimizing the expectation of the poverty function, with a penalty for lifetime ruin.  Finally, in Section \ref{sec3c}, we study properties of the optimal investment strategy.

\subsubsection{Related free-boundary problem}\label{sec3a}

Consider the following FBP on $[0, y_a]$, with $0 < y_d < y_a$, both to be determined.
\begin{align}\label{309}
\left\{\begin{array}{ll}
               \displaystyle \la \hat{L}(y) = - (r-\la) y \hat{L}_y(y) + my^2 \hat{L}_{yy}(y) + (c-A)y + l {\bf 1}_{\{y_d\le y\le y_a\}},
\\ \\
                \hat{L}(0)=0,\qquad \hat{L}_y(y_d)=d,
\\ \\
\hat{L}(y_a)=ay_a+\Pu,\quad  \hat{L}_y(y_a)=a,
              \end{array}
\right.
\end{align}
in which
\begin{align}\notag%\label{310}
m:=\frac{1}{2}\left(\frac{\mu-r}{\sig}\right)^2.
\end{align}

In the following proposition, we present the solution of the FBP \eqref{309}. 

\begin{proposition}\label{prop31}
The solution of the free-boundary problem \eqref{309} on $[0,y_a]$ is given by
\begin{align}\label{311}
\hat{L} (y)=\left\{\begin{array}{ll}
               \displaystyle  k_0y^{\beta_1} + \cAr \, y, \; \, \qquad\quad\qquad\qquad y\in[0,y_d),
\\ \\
                k_1y^{\beta_1}+k_2y^{\beta_2} + \dfrac{c-A}{r} \, y + \dfrac{l}{\la}, \qquad y\in[y_d,y_a], 
          \\
              \end{array}
\right.
\end{align}
in which
\begin{align}\label{312a}
k_0&:= - \cArd \frac{1}{\beta_1} \, y_d^{1-\beta_1},\\\label{312b}
k_1 &:=- \frac{1-\beta_2}{\beta_1-\beta_2}\cAra y_a^{1-\beta_1} - \frac{\beta_2}{\beta_1 - \beta_2} \left(\Pu-\lola\right)y_a^{-\beta_1}, \\\label{312c}
k_2 &:=- \frac{\beta_1-1}{\beta_1-\beta_2}\cAra y_a^{1-\beta_2} + \frac{\beta_1}{\beta_1 - \beta_2} \left(\Pu-\lola\right)y_a^{-\beta_2}, \\\label{312d}
\beta_1 &:= \frac{1}{2m} \left[ (r-\la+m) + \sqrt{(r-\la+m)^2+4\la m} \, \right] > 1, \\\label{312e}
\beta_2 &:= \frac{1}{2m} \left[ (r-\la+m) - \sqrt{(r-\la+m)^2+4\la m} \, \right] < 0.
\end{align}
The ratio of the free boundaries $y_{da}:=\frac{y_d}{y_a}\in(0,1)$ uniquely solves $g(y) = 0$, in which $g$ is defined by
\begin{align}\label{313}
g(y) = & \, \beta_1(1-\beta_2)\cAra \lola \, y^{\beta_1-\beta_2}
+
(\beta_1-\beta_2) \cAra \left(\Pu-\lola\right) y^{\beta_1}\\\notag
&-
(\beta_1-\beta_2) \cArd \lola \, y^{1-\beta_2}
-
(\beta_1-\beta_2) \cArd \left(\Pu - \lola\right) y \\\notag
&+
\beta_2(\beta_1-1) \cAra \lola,
\end{align} 
the free boundary $y_a$ can be expressed in terms of $y_{da}$ via
\begin{align}\label{314}
y_a = \frac{\beta_1}{\beta_1 - 1} \, \frac{\lola \, y_{da}^{-\beta_2} + \Pulola}{\frac{c-A}{r} - a},
\end{align}
and the free boundary $y_d = y_a \cdot y_{da}$.

Moreover, $\hat{L}$ is increasing, concave, and $\calC^2$, except at $y = y_d$, where it is $\calC^1$ and has left- and right-derivatives.
\end{proposition}

\noi \textbf{Proof.}
We start by showing that $y_{da}$ and $y_a$ are well-defined. To this end, we first claim that $g$ in \eqref{313} has a unique zero in $(0, 1)$. Indeed,  
\begin{align}\label{eq:gy}
g_y(y) = (\beta_1-\beta_2) \left[(1-\beta_2) \lola \, y^{-\beta_2} + \Pulola\right] \cdot \left[\beta_1\cAra y^{\beta_1-1} - \cArd \right].
\end{align}
Since $\beta_1>\beta_2$ and $\Pu \ge \lola$, it follows that the first two factors of $g_y$'s expression are positive. The third term changes sign at most once. Therefore, $g$ changes its monotonicity at most once.  Moreover, because $g(0) = \beta_2(\beta_1-1) \cAra \lola < 0$ and $g(1) = (\beta_1-\beta_2) (d - a) \Pu > 0$, it follows that $g$ in \eqref{313} has a unique zero in $(0, 1)$, so $y_{da}$ is well-defined.  Notice also that $y_a > 0$ because $\Pu > \lola$, $\beta_1 > 1$, and $a < \cAr$; thus, $y_d$ is positive and less than $y_a$.

We now turn to showing that the expression given in \eqref{311} is, indeed, the solution of the FBP. First, one can verify that $\hat{L}(y)= j_1y^{\beta_1} + j_2 y^{\beta_2} +\cAr y + \lola {\bf 1}_{\{y_d \le y \le y_a\}}$ solves the differential equation in \eqref{309}, for some constants $j_1$ and $j_2$.  From $\hat{L}(0) = 0$, we deduce that $j_2 = 0$ when $0 \le y < y_a$, and from $\hat{L}_y(y_d-) = d$, we obtain \eqref{312a}.  From the boundary conditions at $y_a$, we obtain \eqref{312b}--\eqref{312e}. 

We now turn to showing that $\hat L$ is $\calC^1$ at $y = d$.  By using the boundary conditions $\hat{L}(y_d+) = \hat{L}(y_d-)$ and $\hat{L}_y(y_d+) = d$, we obtain 
\begin{align}\label{316a}
k_1 &= - \cArd \frac{1}{\beta_1} \, y_d^{1-\beta_1} + \frac{\beta_2}{\beta_1-\beta_2} \, \lola \, y_d^{-\beta_1},\\\label{316b}
k_2 &= -\frac{\beta_1}{\beta_1-\beta_2} \,\lola \, y_d^{-\beta_2}.
\end{align}
By equating \eqref{312b} and \eqref{316a}, we obtain 
\begin{align}\notag%\label{317}
y_{da}^{\beta_1}\left[- \beta_1 (1-\beta_2)\cAra y_a - \beta_1 \beta_2(\beta_1 - \beta_2) \Pulola \right] = - (\beta_1-\beta_2) \cArd y_d + \beta_1 \beta_2 \, \lola,
\end{align}
from which it follows that
\begin{align}\label{318}
y_a = \frac{-\beta_1\beta_2\left[\lola+\Pulola y_{da}^{\beta_1}\right]}{- (\beta_1-\beta_2) \cArd y_{da}+\beta_1(1-\beta_2)\cAra y_{da}^{\beta_1}}.
\end{align}
Similarly, by equating \eqref{312c} and \eqref{316b}, we obtain \eqref{314}.  Finally, by equating the two expressions for $y_a$, \eqref{314} and \eqref{318}, we obtain $g(y_{da})=0$.  In other words, $y_a$ are $y_{da}$ are chosen so that $\hat L$ is $\calC^1$ at $y = d$.

We now show that $\hat{L}$ is increasing and concave on $[0, y_a]$.  Because $\hat{L}_{y}(y_a) = a \ge 0$, it is sufficient to show only concavity, which is accomplished by showing that $\hat{L}_{yy} < 0$ on $[0, y_a]$.  We verify the latter separately on $[0,y_d)$ and on $[y_d,y_a]$. For $y \in [0, y_d)$, we have $\hat{L}_{yy}(y) = - (\beta_1-1) \cArd \frac{1}{y_d}\left(\frac{y}{y_d}\right)^{\beta_1 - 2}$, which is negative because $\beta_1 > 1$ and $d < \cAr$. For $y \in (y_d,y_a]$, algebraic manipulation, together with \eqref{312b}, \eqref{312c}, and \eqref{314}, yields 
\begin{align}\notag%\label{320}
\hat{L}_{yy}(y) \propto & - (1 - \beta_2) \lola \, y_{da}^{-\beta_2} \left( \beta_1 \left( \frac{y}{y_d} \right)^{\beta_1 - \beta_2} - \beta_2 \right) - (\beta_1 - \beta_2) \Pulola   \left( \frac{y}{y_d} \right)^{\beta_1 - \beta_2},
\end{align}
which is negative because $\beta_1>1$, $\beta_2<0$, and $\Pu \ge \lola$.  \qed

\subsubsection{Relation between the free-boundary problem and minimizing poverty with a penalty for ruin}\label{sec3b}

We now show that the Legendre transform of the solution \eqref{311} of the FBP \eqref{309} is the value function $V$ \eqref{305b} and, thereby, provide an implicit expression for $V$.  Because $\hat{L}$ is concave, we can define its convex Legendre transform, as in the following proposition.

\begin{proposition}\label{prop32}
Define the convex Legendre transform of $\hat L$ on $\left[0,\tfrac{c-A}{r} \right]$ by
\begin{align}\label{321}
L(w) := \max_{0\le y\le y_a}\left[\hat{L}(y)-wy\right].
\end{align}
Then, $L$ equals the value function $V$ \eqref{305b} when $l(\cdot)$ is given by \eqref{304b} and $c(\cdot) \equiv c$.
\end{proposition}

\noi \textbf{Proof.}
We show that $L$ satisfies the verification lemma; therefore, $V=L$. Fix $w\in \left[0,\tfrac{c-A}{r} \right]$. The maximizer $y(w)$ solves $\hat{L}_y(y(w))-w=0$. Therefore, $y(w) = (\hat{L}_y)^{-1}(w)=:I(w)$.  (In particular, it follows from \eqref{309} that $y(a) = y_a$ and $y(d) = y_d$.)  More generally, we obtain $L_w(w)=-I(w)=-y(w)$ and $L_{ww}(w)=-1/\hat{L}_{yy}(y(w))$. From these two expressions, it follows that $L$ is decreasing and convex on $\left[0,\tfrac{c-A}{r} \right]$.  

By substituting $y(w) = - L_w(w)$ into the FBP \eqref{309} and by using the expressions obtained in the above paragraph, we deduce that $L$ solves the following BVP on $\left[0,\tfrac{c-A}{r} \right]$.
\begin{align}\label{322}
\left\{\begin{array}{ll}
               \displaystyle  \la L(w) = (rw - (c-A))L_w(w) - m\frac{L_w^2(w)}{L_{ww}(w)} + l {\bf 1}_{\{a\le w\le d\}},
\\ \\
                L(a) = \Pu,\quad L(\tfrac{c-A}{r})=0.
              \end{array}
\right.
\end{align}
The first boundary condition follows from $L(a)=\hat{L}(y_a)-ay_a=\Pu$. To obtain the second boundary condition, notice that because $\hat{L}(y)-\tfrac{c-A}{r} \, y$ is concave and because 
\begin{align}\notag%\label{322b}
\frac{d}{dy}\left.\left[\hat{L}(y) - \cAr \, y\right]\right|_{y=0} = \hat{L}_y(0) - \cAr = 0,
\end{align}
it follows that $\hat{L}(y) - \tfrac{c-A}{r} \, y$ is decreasing on $\left(0,\tfrac{c-A}{r} \right]$.  Because $L\left(\tfrac{c-A}{r} \right) = \max_{0\le y\le y_a}\left[  \hat{L}(y) - \tfrac{c-A}{r} \, y \right]$, it follows that the maximum is attained at $y_{\tfrac{c-A}{r}}=0$; therefore, $L \left(\tfrac{c-A}{r} \right) = 0$.  It is clear that \eqref{322} is equivalent to \eqref{307} when $l(w)$ is given by \eqref{304b} and $c(w) \equiv c$.  \qed

The following theorem provides an implicit expression for the minimum expectation of the poverty function, with a penalty for lifetime ruin, when the consumption rate is constant.

\begin{theorem}\label{thm31}
When $l(\cdot)$ is given by \eqref{304b} and $c(\cdot) \equiv c$, the value function $V$ \eqref{305b} equals
\begin{align}\label{323}
V(w)=\left\{\begin{array}{ll}
               \displaystyle - k_1 (\beta_1 - 1) y^{\beta_1}(w) + k_2 (1 - \beta_2) y^{\beta_2}(w) + \lola,\quad &a\le w\le d,
\\ \\
                \dfrac{\beta_1-1}{\beta_1} \left( \dfrac{c-A}{r} - d \right) y_d\left(\dfrac{c-A-rw}{c-A-rd}\right)^{\frac{\beta_1}{\beta_1-1}},\quad &d<w\le \cAr,
              \end{array}
\right.
\end{align}
and the optimal investment strategy is given in feedback form by $\pi^*_t = \pi^*(W^*_t)$, in which $W^*$ is optimally controlled wealth and $\pi^*$ is defined by
\begin{align}\label{324}
\pi^*(w) = \left\{\begin{array}{ll}
               \displaystyle -\frac{\mu-r}{\sig^2} \left( k_1 \beta_1 (\beta_1 - 1) y^{\beta_1 - 1}(w) - k_2 \beta_2 (1 - \beta_2) y^{\beta_2 - 1}(w) \right), \quad&a\le w < d,
\\ \\
               \dfrac{\mu-r}{\sig^2}(\beta_1-1) \left( \frac{c-A}{r} - w \right) ,\quad &d < w\le \cAr.
              \end{array}
\right.
\end{align}
Here, $k_1$ and $k_2$ are given by \eqref{312b} and \eqref{312c}, respectively, and $y(w) \in [y_d,y_a]$ is the unique solution of 
\begin{align}\label{325}
k_1 \beta_1 y^{\beta_1-1}(w) + k_2 \beta_2 y^{\beta_2-1}(w) + \cAr = w.
\end{align}
\end{theorem}

Notice that, on the interval $\left(d,\tfrac{c-A}{r} \right]$, $\pi^*(w)$ is identical to the optimal strategy in the classical lifetime ruin problem (without poverty); see \cite[Equation (11)]{Young2004}.

\skp \noi \textbf{Proof of Theorem \ref{thm31}.}
From Proposition \ref{prop32}, we know that $V = L$.  Hence, it is sufficient to show that $L(w)$ equals the expression in \eqref{323}.  We separately analyze the two cases $a\le w\le d$ and $d<w\le \tfrac{c-A}{r}$.   If $a\le w\le d$, then, by the boundary conditions in \eqref{309}, 
\begin{align}\label{326}
\frac{d}{dy}\left.\left[\hat{L}(y)-wy\right]\right|_{y=y_d}=\hat{L}_y(y_d)-w=d-w\ge 0,
\end{align}
and
\begin{align}\label{326a}
\frac{d}{dy}\left.\left[\hat{L}(y)-wy\right]\right|_{y=y_a}=\hat{L}_y(y_a)-w=a-w\le 0.
\end{align}
By Proposition \ref{prop31}, $\hat L$ is concave; thus, for a fixed $w$, the function $\hat{L}(y) - wy$ is also concave.  From inequalities \eqref{326} and \eqref{326a}, for which at most one holds with equality, it follows that there exists a unique $y(w) \in [y_d, y_a]$ that maximizes $\hat{L}(y) - wy$, and $y(w)$ can be identified as the unique solution of \eqref{325}. The proof in this case is completed by recalling the definition of $L(w)$ in \eqref{321}.

If $d < w \le \tfrac{c-A}{r}$, then again by the boundary condition in \eqref{309} (because $\hat L$ is $\calC^1$ at $y = y_d$), $\tfrac{d}{dy}\left.[\hat{L}(y)-wy]\right|_{y=y_d}=d-w< 0$.  Again, for a fixed $w$, the function $\hat{L}(y)-wy$ is concave.  So, there exists a unique $y(w) \in [0,y_d)$ that maximizes $\hat{L}(y) - wy$.  Specifically, 
\begin{align}\notag%\label{327}
y(w) = \left(\frac{c - A - rw}{-k_0\beta_1r}\right)^{\frac{1}{\beta_1-1}}.
\end{align}
Then, by using \eqref{312a} and the relationship between $L$ and $\hat L$ in \eqref{321}, the proof is done.  \qed

\subsubsection{Properties of the optimal investment strategy}\label{sec3c}

In this section, we study properties of the optimal investment strategy in \eqref{324}; thus, throughout this section, we assume that $l$ is given by \eqref{304b} and $c(w) \equiv c$.  The following proposition provides necessary and sufficient conditions for the optimal strategy to be monotone as a function of the wealth on each of the intervals $(a, d)$ and $\left(d, \tfrac{c-A}{r} \right)$.

\begin{proposition}
\label{prop34}
Let $\pi^*$ denote the optimal amount invested in the risky asset, as given in \eqref{324}. \medskip \\ \medskip
\indent $($i$)$ $\pi^*$ decreases with $w$ in $\left(d, \tfrac{c-A}{r} \right);$ \\
\smallskip \indent $($ii$)$ $\pi^*$ decreases with $w$ in $(a, d)$ if and only if 
\begin{align}\label{328}
(\beta_1 - 1)(\beta_1-\beta_2)\Pulola y_{da}^{\beta_1} + (1 - \beta_2) \lola \left[ \beta_1(\beta_1 - 1)y_{da}^{\beta_1-\beta_2} + \beta_2(1-\beta_2) \right] \ge 0;
\end{align}
\\
\smallskip \indent $($iii$)$ $\pi^*$ increases with $w$ in $(a, d)$ if and only if
\begin{align}\label{329}
(\beta_1 - 1) \Pulola y_{da}^{\beta_2} + (1 - \beta_2)(\beta_1 + \beta_2 - 1) \lola \le 0.
\end{align}
\smallskip \hangindent 20 pt $($iv$)$ If neither \eqref{328} nor \eqref{329} holds, then there exists $w_0 \in (a, d)$ such that $\pi^*$ decreases with $w$ in $(a, w_0)$ and increases with $w$ in $(w_0, d)$.
\end{proposition}

\smallskip \noi \textbf{Proof.} 
Property (i) immediately follows from the second expression for $\pi^*$ in \eqref{324}.  We now prove properties (ii) - (iv).  %Recall the definition of $y(w)$ from \eqref{325}; then,
%\begin{align}\notag%\label{330}
%\frac{\partial}{\partial w}\pi^*(w)=\frac{\partial}{\partial w}\left[-\frac{\mu-r}{\sig^2}\hat{L}_{yy}(y(w))\cdot y(w)\right] = -\frac{\mu-r}{\sig^2}\frac{\partial}{\partial y}\left[\hat{L}_{yy}(y(w))\cdot y(w)\right]\frac{\partial}{\partial w}y(w).
%\end{align}
First, because $y(w) = - V_w(w)$ and because $V$ is convex, it follows that $\frac{\partial}{\partial w}y(w) < 0$.  Thus,
\begin{align*}
&\frac{\partial}{\partial w}\pi^*(w) \propto \frac{\partial}{\partial y} \left( k_1 \beta_1 (\beta_1 - 1) y^{\beta_1 - 1} - k_2 \beta_2 (1 - \beta_2) y^{\beta_2 - 1} \right) \\
&\quad \propto k_1 \beta_1 (\beta_1 - 1)^2 y^{\beta_1 - \beta_2} + k_2 \beta_2 (1 - \beta_2)^2 \\
&\quad \propto - (\beta_1 - 1) \left[ \beta_1(1 - \beta_2) \lola + (\beta_1 - \beta_2) \Pulola y_{da}^{\beta_2} \right] \left( \frac{y}{y_a} \right)^{\beta_1 - \beta_2} - \beta_2(1 - \beta_2)^2 \lola =: p(y).
\end{align*}
Note that $p(y)$ decreases with $y$.  Thus, $p(y) < 0$ for all $y \in (y_d, y_a)$ if and only if $p(y_d) \le 0$, which is equivalent to inequality \eqref{328}.  Similarly, $p(y) > 0$ for all $y \in (y_d, y_a)$ if and only if $p(y_a) \ge 0$, which is equivalent to inequality \eqref{329}.  Finally, if neither \eqref{328} nor \eqref{329} holds, then there exists $y_0 \in (y_d, y_a)$ such that $p(y) > 0$ for $y \in (y_d, y_0)$ and $p(y) < 0$ for $y \in (y_0, y_a)$.  By setting $w_0$ equal to the expression on the left side of \eqref{325}, item (iv) follows.   \qed

%From \eqref{325}, \eqref{316a}, \eqref{316b}, and since $\beta_1>1,\beta_2<0$ it follows that
%\begin{align}\notag%\label{331}
%\frac{\partial}{\partial w}y(w)=\frac{1}{k_1\beta_1(\beta_1-1)y^{\beta_1-2}(w)+k_2\beta_2(\beta_2-1)y^{\beta_2-2}(w)}<0.
%\end{align}
%Therefore, the sign of $\tfrac{\partial}{\partial w}\pi^*(w)$ is the same as the sign of $\tfrac{\partial}{\partial y}[\hat{L}_{yy}(y)\cdot y]$. Now, 
%\begin{align}\label{332}
%&\frac{\partial}{\partial y}\left[\hat{L}_{yy}(y)\cdot y\right]=-\frac{(\beta_1-1)\cAra}{y_a(\lola+\Pulola y_{da}^{\beta_2})}\left(\frac{y}{y_a}\right)^{\beta_2-1}\cdot p(y),
%\end{align}
%where
%\begin{align}\notag%\label{333}
% p(y)
%=\beta_1\left(\beta_1(1-\beta_2)\lola+(\beta_1-\beta_2)\Pulola y_{da}^{\beta_2}\right)\left(\frac{y}{y_a}\right)^{\beta_1-\beta_2}-\beta_2^2(1-\beta_2)\lola.
%\end{align}
%The coefficient of $p(y)$ in \eqref{332} is negative. Hence, we analyze $p(y)$. Since $\beta_1>0>\beta_2$ and $\Pu>\lola$ we get that 
%\begin{align}\notag%\label{334}
% p'(y)
%=\beta_1(\beta_1-\beta_2)\left(\beta_1(1-\beta_2)\lola+(\beta_1-\beta_2)\Pulola y_{da}^{\beta_2}\right)\left(\frac{y}{y_a}\right)^{\beta_1-\beta_2-1}-\beta_2^2(1-\beta_2)\lola>0.
%\end{align}
%Hence if $p(y_d)\ge 0$ (resp., $p(y_a)\le 0$) then $p(y)>0$ on $(y_d,y_a]$ (resp., $p(y)<0$ on $[y_d,y_a)$) and so $\tfrac{\partial}{\partial y}[\hat{L}_{yy}(y)\cdot y]<0$ (resp., $\tfrac{\partial}{\partial y}[\hat{L}_{yy}(y)\cdot y]>0$) on the same interval, which is equivalent to condition (ii) (resp., (iii)).  

Next, we examine how $\pi^*$ varies with some parameters of the model.  To that end, we begin with a lemma concerning $y_{da}$.

\begin{lemma}\label{lem:yda}
The ratio of the free boundaries, $y_{da}$, satisfies the following inequality.
\begin{equation}\label{ineq:yda}
y_{da} > \left( \frac{{c-A} - rd}{{c-A} - ra} \right)^{\frac{1}{\beta_1 - 1}},
\end{equation}
from which it follows that $y_{da}$ increases with $l$ and decreases with $\Pu$.
\end{lemma}

\noi \textbf{Proof.} 
Because $y_{da}$ is the unique zero in $(0, 1)$ of $g$ given by \eqref{313}, and because $g(0) < 0$ and $g(1) > 0$, it follows that inequality \eqref{ineq:yda} holds if and only if
\begin{equation*}
g \left( \left( \frac{{c-A} - rd}{{c-A} - ra} \right)^{\frac{1}{\beta_1 - 1}} \right) < 0,
\end{equation*}
which is equivalent to
\begin{equation*}
\left( \frac{{c-A} - rd}{{c-A} - ra} \right)^{\frac{\beta_1 - \beta_2}{\beta_1 - 1}} < 1,
\end{equation*}
which is true.  Thus, we have proved inequality \eqref{ineq:yda}.

Differentiate $g(y_{da}) = 0$ fully with respect to $l$ to obtain
\begin{equation}\label{eq:gyyda}
0 = g_y(y_{da}) \, \frac{\partial y_{da}}{\partial l} + \left. \frac{\partial g(y)}{\partial l} \right|_{y = y_{da}};
\end{equation}
thus, because $g_y(y_{da}) > 0$, to show that $y_{da}$ increases with $l$, it is enough to show that  
\begin{equation*}
\left. \frac{\partial g(y)}{\partial l} \right|_{y = y_{da}} < 0.
\end{equation*}
It is straightforward to show that
\begin{equation}\label{eq:glyda}
\left. \frac{\partial g(y)}{\partial l} \right|_{y = y_{da}} = - \frac{1}{l} (\beta_1 - \beta_2) \Pu \left\{ \cAra y_{da}^{\beta_1} - \cArd y_{da} \right\},
\end{equation}
which is negative due to inequality \eqref{ineq:yda}.

The proof that $y_{da}$ decreases with $\Pu$ is similar, so we omit it.  \qed

As the penalty for being in poverty $l$ increases relative to the penalty for ruin $\Pu$, we expect the optimal investment strategy to increase with $l$ (and decrease with $\Pu$) because the individual has more incentive to get out of poverty, and investing more heavily in the risky asset is the best way to do that.  (Recall that we assume $\Pu \ge \lola$ so that the individual will not wish to commit financial suicide.)

\begin{proposition}\label{prop:pi_incr_ell}
The optimal amount to invest in the risky asset $\pi^*(w)$ $($weakly$)$ increases with $l$ and $($weakly$)$ decreases with $\Pu$ for $w \in (a, d)$ and is independent of $l$ and $\Pu$ for $w \in \left( d, \frac{c -A}{r} \right)$.
\end{proposition}

\noi \textbf{Proof.} 
It is clear, from the expression for $\pi^*$ in \eqref{324} on $\left( d, \frac{c -A}{r} \right)$, that $\pi^*$ is independent of $l$ and $\Pu$ on this interval. Thus, we focus on showing that $\pi^*(w)$ increases with $l$ for $w \in (a, d)$.  First, we prove that $\pi^*(a+)$ increases with $l$.  By using equations \eqref{312b}, \eqref{312c}, and \eqref{314}, we get
\begin{equation}\label{eq:pia}
\pi^*(a+) = \frac{\mu - r}{\sig^2}(\beta_1 - 1) \cAra \frac{(1 - \beta_2) \lola \, y_{da}^{-\beta_2} + \Pulola}{ \lola \, y_{da}^{-\beta_2} + \Pulola}.
\end{equation}
Differentiate this expression with respect to $l$ to obtain
\begin{equation*}
\frac{\partial \pi^*(a+)}{\partial l} \propto \Pu - \beta_2 \, l \Pulola \, \frac{1}{y_{da}} \, \frac{\partial y_{da}}{\partial l},
\end{equation*}
which is positive because $\beta_2 < 0$ and $y_{da}$ increases with $l$ from Lemma \ref{lem:yda}.  Thus, $\pi^*(a+)$ increases with $l$.

Next, we prove that $\pi^*(d-)$ increases with $l$.  By using equations \eqref{316a}, \eqref{316b}, and \eqref{314}, we get
\begin{equation}\label{eq:pid}
\pi^*(d-) = \frac{\mu - r}{\sig^2} \, (\beta_1 - 1) \left[ \cArd - \beta_2 \cAra \frac{\lola}{\lola y_{da}^{1 - \beta_2} + \Pulola y_{da}}\right].
\end{equation}
Differentiate this expression with respect to $l$ to obtain
\begin{align*}
\frac{\partial \pi^*(d-)}{\partial l} &\propto \Pu \, y_{da} - l \left( (1 - \beta_2) \lola \, y_{da}^{-\beta_2} + \Pulola \right) \frac{\partial y_{da}}{\partial l} \\
&\propto (\beta_1 - 1) \cAra y_{da}^{\beta_1 - 1} > 0,
\end{align*}
in which the second line follows from \eqref{eq:gyyda}, \eqref{eq:glyda}, and \eqref{eq:gy}.  Thus, $\pi^*(d-)$ increases with $l$.

Next, we derive a differential equation for $\pi^*$ on $(a, d)$ that we use to show that $\pi^*$ (weakly) increases with $l$ on that interval.  Rewrite the differential equation in \eqref{322} as follows; we write $V$ in place of $L$.
\begin{equation*}
\la V = \left[ (rw - c + A) + \frac{\mu - r}{2} \, \pi^* \right] V_w + l.
\end{equation*}
Differentiate this expression with respect to $w$, divide both sides by $V_w$, and rearrange the result to obtain
\begin{equation}\label{eq:piODE}
\frac{\mu - r}{2} \, \pi^*_w  = \la - r + m + \frac{\mu - r}{\sig^2} \, \frac{rw - c + A}{\pi^*}.
\end{equation}

Finally, suppose $l_1 \le l_2$, and let $\pi_i(\cdot) = \pi^*(\cdot; l = l_i)$ for $i = 1, 2$.  We wish to show that $\pi_1(w) \le \pi_2(w)$ for all $w \in (a, d)$.  Suppose, on the contrary, that $\pi_1(w) > \pi_2(w)$ for some $w \in (a, d)$; then, because $\pi_1(a+) < \pi_2(a+)$ and $\pi_1(d-) < \pi_2(d-)$, it follows that $\pi_1 - \pi_2$ takes a positive maximum at some $w_0 \in (a, d)$.  Then,
\begin{align*}
0 &= \frac{\mu - r}{2} \left( (\pi_1)_w(w_0) - (\pi_2)_w(w_0) \right) \\
&= \frac{\mu - r}{\sig^2} (r w_0 - c + A) \left( \frac{1}{\pi_1(w_0)} - \frac{1}{\pi_2(w_0)} \right) > 0,
\end{align*}
a contradiction.  Thus, $\pi_1 \le \pi_2$ in $(a, d)$.

A similar proof shows that $\pi^*(w)$ (weakly) decreases with $\Pu$ for $w \in (a, d)$, so we omit that proof.  \qed

In a related result, we show that as $l \to 0+$, $\pi^*$ approaches the optimal investment strategy for minimizing the probability of lifetime ruin.

\begin{proposition}\label{prop:pi_cont_ell}
\begin{equation*}
\lim_{l \to 0+} \pi^*(w) = \pi^0(w),
\end{equation*}
for all $w \in \left(a, \frac{c-A}{r} \right)$,  in which $\pi^0(w) := \frac{\mu - r}{\sig^2} \, (\beta_1 - 1) \left( \frac{c-A}{r} - w \right)$ is the optimal investment strategy for minimizing the probability of lifetime ruin when $c(w) \equiv c$.
\end{proposition}

\noi \textbf{Proof.} The result is clear for $w \in \left(d, \frac{c-A}{r} \right)$ because $\pi^*(w) = \pi^0(w)$ for all $w \in \left(d, \frac{c-A}{r} \right)$.  For $w \in (a, d)$, $\pi^*$ is determined by its value at $w = a+$ and its differential equation in \eqref{eq:piODE}.  Note that, in the differential equation, $\pi^*_w$ depends only on the parameter $l$ via $\pi^*$.  Thus, it is enough to show that $\lim_{l \to 0+} \pi^*(a+) = \pi^0(a+)$, which is clear from \eqref{eq:pia} because $\lim_{l \to 0+} y_{da}^{-\beta_2} \in [0, 1]$.  \qed

\begin{remark}
Note that inequality \eqref{328} strictly holds when $l = 0$; thus, when $l$ is small enough, $\pi^*$ decreases with $w$ on $(a, d)$.  In other words, when $l$ is small enough, $\pi^*$ behaves like $\pi^0$ on $(a, d)$ in that $\pi^*$ decreases with $w$ and is ``close to'' $\pi^0$.
\end{remark}

As a corollary to Propositions \ref{prop:pi_incr_ell} and \ref{prop:pi_cont_ell}, we observe that the optimal amount invested in the risky asset is greater in our model than when simply minimizing the probability of lifetime ruin.

\begin{corollary}\label{pi0}
For $a < w < d$,
$$
\pi^*(w) > \pi^0(w),
$$
and for $d < w < \cAr$,
$$
\pi^*(w) = \pi^0(w).
$$
In particular, $\pi^*(d-) > \pi^*(d+)$.
\end{corollary}

\noi \textbf{Proof.} This result follows immediately from Propositions \ref{prop:pi_incr_ell} and \ref{prop:pi_cont_ell} because $\pi^0$ is the optimal strategy when the penalty for poverty $l = 0$. \qed

\begin{remark}
\label{rem:scar}
Because $\pi^*(w) = \pi^0(w)$ for $w > d$, the individual is myopic in her investment when she is not in poverty.  That is, she invests as if her only penalty is $\Pu$ when wealth reaches the ruin level $a$.  In their conclusion, Mullainathan and Shafir \cite{MullShafir2013} hypothesize that the seeds of scarcity exists when resources are abundant ($w > d$ for our problem) in that people myopically waste their resources and scarcity occurs more quickly than if the people had used their resources more effectively when they were abundant.

On the other hand, $\pi^0$ is independent of the ruin level; thus, one could also say that our individual invests as if she were going to receive the full penalty when wealth reaches $d > a$, the poverty level.
\end{remark}

\subsection{Proportional rate of consumption}\label{sec4}

In this section, we solve the optimization problem when the rate of consumption of proportional to wealth, that is, $c(w) = \kap w$.   We assume that $\kap > r$ and that $a > \frac{A}{\kap-r}$ in order to avoid a trivial case; see Section 4.1 in \cite{Young2004}.  In the case of proportional consumption, there is no safe level; more precisely, $\ws = \infty$.

The analysis is similar to the one presented in Section \ref{sec30}.  Specifically, we first study an auxiliary FBP and then connect it to the the problem of minimizing the expectation of the poverty function, with a penalty for lifetime ruin, via the Legendre transform.  Finally, we study properties of the optimal investment strategy.  Whenever the proofs are similar to the corresponding ones in Section \ref{sec30}, we omit them.

\subsubsection{Related free-boundary problem}\label{sec4a}

Consider the following FBP on $[0, z_a]$ with $0 < z_d < z_a$, both to be determined.
\begin{align}\label{401}
\left\{\begin{array}{ll}
               \displaystyle \la \hat{M}(z) = -(r - \kap - \la) z \hat{M}(z) +  mz^2\hat{M}_{zz}(z) - Az + l {\bf 1}_{\{z_d \le z \le z_a\}},
\\ \\
                \hat{M}(0) = 0,\quad \hat{M}_z(z_d) = d,
\\ \\
\hat{M}(z_a) = az_a + \Pu,\quad \hat{M}_z(z_a) = a,
              \end{array}
\right.
\end{align}
in which, as before,
\begin{align}\notag%\label{401b}
m = \frac{1}{2}\left(\frac{\mu-r}{\sig}\right)^2.
\end{align}

In the following proposition, we present the solution of the FBP \eqref{401}. 
\begin{proposition}\label{prop41}
The solution of the FBP \eqref{401} on $[0, z_a]$ is given by
\begin{align}\label{402}
\hat{M} (z)=\left\{\begin{array}{ll}
               \displaystyle  k_4z^{\gam_1} + \Akr \, z,\qquad\quad\qquad &z\in[0, z_d),
\\ \\
                k_5z^{\gam_1}+k_6z^{\gam_2}+ \dfrac{A}{\kap - r} \, z + \dfrac{l}{\la}, \quad &z\in[z_d, z_a], 
              \end{array}
\right.
\end{align}
in which
\begin{align}\notag%\label{403a}
k_4&:= \dAkr \frac{1}{\gam_1} \, z_d^{1-\gam_1},\\\label{403b}
k_5 &:= \frac{1-\gam_2}{\gam_1-\gam_2} \aAkr z_a^{1-\gam_1}-\frac{\gam_2}{\gam_1-\gam_2}\left(\Pu-\lola\right)z_a^{-\gam_1},\\\label{403c}
k_6 &:= - \frac{1-\gam_1}{\gam_1-\gam_2} \aAkr z_a^{1-\gam_2}+\frac{\gam_1}{\gam_1-\gam_2}\left(\Pu-\lola\right)z_a^{-\gam_2},\\\notag
\gam_1& := \frac{1}{2m} \left[(r-\kap-\la+m) + \sqrt{(r-\kap- \la+m)^2+4\la m} \, \right] \in (0,1),\\\notag
\gam_2&:=  \frac{1}{2m} \left[(r-\kap-\la+m) - \sqrt{(r-\kap- \la+m)^2+4\la m} \, \right] < 0.\\\notag
\end{align}
The ratio of the free boundaries $z_{da} := \frac{z_d}{z_a} \in (0,1)$ uniquely solves $h(z) = 0$, in which $h$ is defined by
\begin{align}\label{404}
h(z) = & \, - \gam_1(1-\gam_2) \aAkr \lola z^{\gam_1-\gam_2}
- (\gam_1-\gam_2) \aAkr \left(\Pu-\lola\right) z^{\gam_1}\\\notag
&+ (\gam_1-\gam_2) \left(d - \Akr \right) \lola z^{1-\gam_2}
+ (\gam_1-\gam_2) \left(d-\Akr\right) \left(\Pu-\lola\right) z \\\notag
&- \gam_2(\gam_1-1) \aAkr \lola,
\end{align}
the free boundary $z_a$ can be expressed in terms of $z_{da}$ via
\begin{align}\label{405}
z_a=\frac{\gam_1}{1-\gam_1}\frac{\lola \, z_{da}^{-\gam_2} + \Pulola}{a-\frac{A}{\kap-r}},
\end{align}
and the free boundary $z_d = z_a \cdot z_{da}$.

Moreover, $\hat{M}$ is increasing, concave, and $\mathcal{C}^2$, except at $z=z_d$, where it is $\mathcal{C}^1$ and has left- and right-derivatives.
\end{proposition}

\noi \textbf{Proof.}
The proof is similar to the proof of Proposition \ref{prop31}. The parameters $z_{da}$ and $z_a$ are well-defined. Indeed, 
\begin{align}\label{eq:hz}
h_z(z)=(\gam_1-\gam_2)\left[(1-\gam_2)\lola z^{-\gam_2} + \Pulola\right]\cdot\left[ -\gam_1\aAkr z^{\gam_1-1}+\dAkr\right].
\end{align}
Since $\gam_1>\gam_2$ and $\Pu \ge \lola$, it follows that the first two factors of $h_z$'s expression are positive. The third term changes sign at most once. Therefore, $h$ changes its monotonicity at most once.  Moreover, because $h(0) = \gam_2(1 - \gam_1)\aAkr \lola < 0$ and $h(1) = (\gam_1-\gam_2)\Pu(d-a)>0$, it follows that $h$ in \eqref{404} has a unique zero in $(0, 1)$, so $z_{da}$ is well-defined.  Notice also that $z_a > 0$ because $\Pu > \lola$, $\gam_1 < 1$, and $a > \Akr$; thus, $z_d$ is positive and less than $z_a$. 

The argument that the expression in \eqref{402} (with the above defined $z_a$ and $z_d$) solves the FBP is almost identical to the one in the proof of Proposition \ref{prop31}; therefore, we omit it.

The proof that $\hat{M}$ is increasing and concave is also similar. The main difference is in the explicit expression of $\hat{M}_{zz}$. For $z\in[0, z_d)$, we have $\hat{M}_{zz}(z) = - (1 - \gam_1)\dAkr\frac{1}{z_d}\left(\frac{z}{z_d}\right)^{\gam_1}$, which is negative because $\gam_1<1$ and $d > \Akr$. For $z \in (z_d, z_A]$, algebraic manipulation, together with \eqref{403b}, \eqref{403c}, and \eqref{405}, yields
\begin{align}\notag%\label{407}
\hat{M}_{zz}(z) \propto & - (1 - \gam_2) \lola \, z_{da}^{-\gam_2} \left( \gam_1 \left( \frac{z}{z_d} \right)^{\gam_1 - \gam_2} - \gam_2 \right) - (\gam_1 - \gam_2) \Pulola   \left( \frac{z}{z_d} \right)^{\gam_1 - \gam_2},
\end{align}
which is negative because $\gam_1 > 0$, $\gam_2 < 0$, and $\Pu \ge \lola$.  \qed

\subsubsection{Relation between the free-boundary problem and minimizing poverty with a penalty for ruin}\label{sec4b}

In this section, we show that the Legendre transform of the solution \eqref{402} of the FBP \eqref{401} is the value function given in \eqref{305b} and, thereby, provide an implicit expression for it.  Because $\hat{M}$ is concave, we can define its convex Legendre transform, as in the following proposition.

\begin{proposition}\label{prop42}
Define the convex Legendre transform of $\hat M$ on $[0, \infty)$ by
\begin{align}\notag%\label{408}
M(w) := \max_{0\le z \le z_a}\left[\hat{M}(z) - wz \right].
\end{align}
Then, $M$ equals the value function $V$ \eqref{305b} when $l(\cdot)$ is given by \eqref{304b} and $c(w) = \kap w$.
\end{proposition}

\noi \textbf{Proof.} As in the proof of Proposition \ref{prop32}, the conclusion that $V = M$ follows by showing that $M$ satisfies the conditions of the verification lemma. The main difference between the proofs lies in the boundary condition at $w = \ws$.  Specifically, we need to show that $\lim_{w\to\infty} M(w) = 0$.  For $w < d$, we have 
\begin{align}\notag%\label{408b}
\frac{d}{dz} \left. \left[\hat{M}(z) - wz \right] \right|_{z=z_d} = \hat{M}_z(z_d) - w = d - w < 0, 
\end{align}
so the $z(w)$ that  maximizes $\hat{M}(z)-wz$ lies in $(0,z_d)$, and satisfies $\hat{M}_z(z(w)) - w = 0$. It follows that $z(w) = \left(\frac{w-\Akr}{k_4 \gam_1} \right)^{-\frac{1}{1 - \gam_1}}$, and $M(w) =  \hat{M}(z(w)) - wz(w) = k_4 (1 - \gam_1) \left( \frac{w - \Akr}{k_4 \gam_1} \right)^{- \frac{\gam_1}{1 - \gam_1}}$.  Thus, $\lim_{w\to\infty} M(w) = 0$, as required.  \qed

The following theorem provides an implicit expression for the value function of the minimum expectation of the poverty function, with a penalty for lifetime ruin, when the consumption rate is proportional to wealth.  The proof is similar to the one of Theorem \ref{thm31}, so we omit it.

\begin{theorem}\label{thm42}
When $l(\cdot)$ is given by \eqref{304b} and $c(w) = \kap w$, the value function $V$ \eqref{305b} equals 
\begin{align}\notag%\label{409}
V(w)=\left\{\begin{array}{ll}\notag
               \displaystyle k_5(1-\gam_1)z^{\gam_1}(w)+k_6(1-\gam_2)z^{\gam_2}(w)+\lola,\quad &a \le w \le d,
\\ \\
                \dfrac{1-\gam_1}{\gam_1}\left( d - \dfrac{A}{\kap - r} \right) z_d\left(\frac{w-\Akr}{d-\Akr}\right)^{-\frac{\gam_1}{1-\gam_1}},\quad &w > d,
              \end{array}
\right.
\end{align}
and the optimal investment strategy is given in feedback form by $\pi^*_t = \pi^*(W^*_t)$, in which $W^*$ is optimally controlled wealth and $\pi^*$ is defined by
\begin{align}\label{410}
\pi^*(w) = \left\{\begin{array}{ll}
               \displaystyle \frac{\mu-r}{\sig^2}\left(k_5\gam_1 (1-\gam_1)z^{\gam_1-1}(w)+k_6\gam_2(1-\gam_2) z^{\gam_2-1}(w)\right), \quad&a\le w < d,
\\ \\
               \dfrac{\mu-r}{\sig^2}(1-\gam_1) \left( w - \frac{A}{\kap- r} \right) ,\quad &w > d.
              \end{array}
\right.
\end{align}
Here, $k_5$ and $k_6$ are given by \eqref{403b} and \eqref{403c}, respectively, and $z(w) \in [z_d, z_a]$ is the unique solution of 
\begin{align}\notag%\label{411}
k_5 \gam_1 z^{\gam_1-1}(w) + k_6 \gam_2 z^{\gam_2-1}(w) + \Akr = w.
\end{align}
\end{theorem}

\subsubsection{Properties of the optimal investment strategy}\label{sec4c}

In this section, we study properties of the optimal investment strategy in \eqref{410}; thus, throughout this section, we assume that $l(\cdot)$ is given by \eqref{304b} and $c(w) = \kap w$. The following proposition states that the strategy is increasing as a function of the wealth on each of the intervals $(a, d)$ and $(d,\infty)$.

\begin{proposition}\label{prop44}
The optimal amount invested in the risky asset, as given in \eqref{410}, increases with $w$ in $(a, d)$ and in $(d, \infty)$.
\end{proposition}

\smallskip \noi \textbf{Proof.} 
That $\pi^*$ increases with $w$ in $(d, \infty)$ immediately follows from the second expression for $\pi^*(w)$ in \eqref{410}.  Now, consider $w \in (a, d)$.  As in the proof of Proposition \ref{prop34}, we observe that, because $z(w) = - V_w(w)$ and because $V$ is convex, it follows that $\frac{\partial}{\partial w}z(w) < 0$.  Thus,
\begin{align*}
&\frac{\partial}{\partial w}\pi^*(w) \propto \frac{\partial}{\partial z} \left( - k_5 \gam_1 (1 - \gam_1) z^{\gam_1 - 1} - k_6 \gam_2 (1 - \gam_2) z^{\gam_2 - 1} \right) \\
&\quad \propto k_5 \gam_1 (1 - \gam_1)^2 z^{\gam_1 - \gam_2} + k_6 \gam_2 (1 - \gam_2)^2 \\
&\quad \propto (1 - \gam_1) \left[ \gam_1(1 - \gam_2) \lola + (\gam_1 - \gam_2) \Pulola z_{da}^{\gam_2} \right] \left( \frac{z}{z_a} \right)^{\gam_1 - \gam_2} - \gam_2(1 - \gam_2)^2 \lola > 0.
\end{align*}
\qed

Next, we examine how $\pi^*$ varies with some parameters of the model.  To that end, we begin with a lemma concerning $z_{da}$.

\begin{lemma}\label{lem:zda}
The ratio of the free boundaries, $z_{da}$, satisfies the following inequality.
\begin{equation}\label{ineq:zda}
z_{da} > \left( \frac{{d(\kappa-r)} - A}{{a(\kappa-r)} -A} \right)^{\frac{1}{\gam_1-1}},
\end{equation}
from which it follows that $z_{da}$ increases with $l$ and decreases with $\Pu$.
\end{lemma}

\noi \textbf{Proof.} 
Because $z_{da}$ is the unique zero in $(0, 1)$ of $h$ given by \eqref{404}, and because $h(0) < 0$ and $h(1) > 0$, it follows that inequality \eqref{ineq:zda} holds if and only if
\begin{equation*}
h \left( \left(\frac{{d(\kappa-r)} - A}{{a(\kappa-r)} -A}  \right)^{\frac{1}{\gam_1-1}} \right) > 0,
\end{equation*}
which is equivalent to
\begin{equation*}
\left( \frac{{d(\kappa-r)} - A}{{a(\kappa-r)} -A} \right)^{\frac{\gam_1 - \gam_2}{\gam_1-1}} < 1,
\end{equation*}
which is true.  Thus, we have proved inequality \eqref{ineq:zda}.

Differentiate $h(z_{da}) = 0$ fully with respect to $l$ to obtain
\begin{equation}\label{eq:hzzda}
0 = h_z(z_{da}) \, \frac{\partial z_{da}}{\partial l} + \left. \frac{\partial h(z)}{\partial l} \right|_{z = z_{da}};
\end{equation}
thus, because $h_z(z_{da}) > 0$, to show that $z_{da}$ increases with $l$, it is enough to show that  
\begin{equation*}
\left. \frac{\partial h(z)}{\partial l} \right|_{z = z_{da}} < 0.
\end{equation*}
It is straightforward to show that
\begin{equation}\label{eq:hlzda}
\left. \frac{\partial h(z)}{\partial l} \right|_{z = z_{da}} = \frac{1}{l} (\gam_1 - \gam_2) \Pu \left\{  \aAkr z_{da}^{\gam_1} - \dAkr z_{da} \right\},
\end{equation}
which is negative due to inequality \eqref{ineq:zda}.

The proof that $z_{da}$ decreases with $\Pu$ is similar, so we omit it.  \qed

As the penalty for being in poverty $l$ increases relative to the penalty for ruin $\Pu$, we expect the optimal investment strategy to increase with $l$ (and decrease with $\Pu$) because the individual has more incentive to get out of poverty, as in Section \ref{sec3c}.

\begin{proposition}\label{prop:pi_incr_ell2}
The optimal amount to invest in the risky asset $\pi^*(w)$ $($weakly$)$ increases with $l$ and $($weakly$)$ decreases with $\Pu$ for $w \in (a, d)$ and is independent of $l$ and $\Pu$ for $w \in \left( d, \infty \right)$.
\end{proposition}

\noi \textbf{Proof.} 
It is clear, from the expression for $\pi^*$ in \eqref{410} on $\left( d,\infty \right)$, that $\pi^*$ is independent of $l$ and $\Pu$ on this interval. Thus, we focus on showing that $\pi^*(w)$ increases with $l$ for $w \in (a, d)$.  First, we prove that $\pi^*(a+)$ increases with $l$.  By using equations \eqref{403b}, \eqref{403c}, and \eqref{405}, we get 
\begin{equation}
\label{eq:pia2}
\pi^*(a+) = \frac{\mu - r}{\sig^2}(1-\gam_1) \aAkr \frac{(1 - \gam_2) \lola \, z_{da}^{-\gam_2} + \Pulola}{ \lola \, z_{da}^{-\gam_2} + \Pulola}.
\end{equation}
Differentiate this expression with respect to $l$ to obtain
\begin{equation*}
\frac{\partial \pi^*(a+)}{\partial l} \propto \Pu - \gam_2 \, l \Pulola \, \frac{1}{z_{da}} \, \frac{\partial z_{da}}{\partial l},
\end{equation*}
which is positive because $\gam_2 < 0$ and $z_{da}$ increases with $l$ from Lemma \ref{lem:zda}.  Thus, $\pi^*(a+)$ increases with $l$.

Next, we prove that $\pi^*(d-)$ increases with $l$.  By finding expressions for $k_5$ and $k_6$ analogous to equations \eqref{316a} and \eqref{316b}, and by \eqref{405}, we get
\begin{equation}
\label{eq:pid2}
\pi^*(d-) = \frac{\mu - r}{\sig^2} \, (1 - \gam_1) \left[ \dAkr - \gam_2 \aAkr \frac{\lola}{\lola z_{da}^{1 - \gam_2} + \Pulola z_{da}}\right].
\end{equation}
Differentiate this expression with respect to $l$ to obtain
\begin{align*}
\frac{\partial \pi^*(d-)}{\partial l} &\propto \Pu \, z_{da} - l \left( (1 - \gam_2) \lola \, z_{da}^{-\gam_2} + \Pulola \right) \frac{\partial z_{da}}{\partial l} \\
&\propto \frac{(1 - \gam_1) \aAkr z_{da}^{ 1-\gam_1} }{- \gam_1\aAkr z_{da}^{\gam_1-1} + \dAkr} > 0,
\end{align*}
in which the second line follows from \eqref{eq:hzzda}, \eqref{eq:hlzda}, and \eqref{eq:hz}. The inequality follows since both the numerator and denominator are positive, in which the latter follows from $h_z(z_{da}) > 0$; see the arguments in the paragraph that follows \eqref{eq:hz}. Thus, $\pi^*(d-)$ increases with $l$.

Parallel to \eqref{eq:piODE}, we have the following differential equation for $\pi^*$ on $(a, d)$ that we use to show that $\pi^*$ (weakly) increases with $l$ on that interval. 
\begin{equation}
\label{eq:piODE2}
\frac{\mu - r}{2} \, \pi^*_w  = \la - r + \kap + m + \frac{\mu - r}{\sig^2} \, \frac{(r-\kappa)w + A}{\pi^*}.
\end{equation}
Suppose $l_1 \le l_2$, and let $\pi_i(\cdot) = \pi^*(\cdot; l = l_i)$ for $i = 1, 2$.  We wish to show that $\pi_1(w) \le \pi_2(w)$ for all $w \in (a, d)$.  Suppose, on the contrary, that $\pi_1(w) > \pi_2(w)$ for some $w \in (a, d)$; then, because $\pi_1(a+) < \pi_2(a+)$ and $\pi_1(d-) < \pi_2(d-)$, it follows that $\pi_1 - \pi_2$ takes a positive maximum at some $w_0 \in (a, d)$.  Then,
\begin{align*}
0 &= \frac{\mu - r}{2} \left( (\pi_1)_w(w_0) - (\pi_2)_w(w_0) \right) \\
&= \frac{\mu - r}{\sig^2} ((r-\kappa) w_0 + A) \left( \frac{1}{\pi_1(w_0)} - \frac{1}{\pi_2(w_0)} \right) > 0,
\end{align*}
a contradiction.  Thus, $\pi_1 \le \pi_2$ in $(a, d)$.

A similar proof shows that $\pi^*(w)$ (weakly) decreases with $\Pu$ for $w \in (a, d)$, so we omit that proof.  \qed

In a related result, we show that as $l \to 0+$, $\pi^*$ approaches the optimal investment strategy for minimizing the probability of lifetime ruin.

\begin{proposition}\label{prop:pi_cont_ell2}
\begin{equation*}
\lim_{l \to 0+} \pi^*(w) = \pi^0(w),
\end{equation*}
for all $w \in \left(a, \infty \right)$,  in which $\pi^0(w) := \frac{\mu - r}{\sig^2} \, (1-\gam_1) \wAkr $ is the optimal investment strategy for minimizing the probability of lifetime ruin when $c(w) \equiv \kappa w$.
\end{proposition}

\noi \textbf{Proof.} The result is clear for $w \in \left(d, \infty \right)$ because $\pi^*(w) = \pi^0(w)$ for all $w \in \left(d, \infty \right)$.  For $w \in (a, d)$, $\pi^*$ is determined by its value at $w = a+$ and its differential equation in \eqref{eq:piODE2}.  Note that, in the differential equation, $\pi^*_w$ depends only on the parameter $l$ via $\pi^*$.  Thus, it is enough to show that $\lim_{l \to 0+} \pi^*(a+) = \pi^0(a+)$, which is clear from \eqref{eq:pia2} because $\lim_{l \to 0+} y_{da}^{-\beta_2} \in [0, 1]$.  \qed

As a corollary to Propositions \ref{prop:pi_incr_ell2} and \ref{prop:pi_cont_ell2}, parallel to Corollary \ref{pi0}, we observe that $\pi^*(w) \ge \pi^0(w)$.  The comments in Remark \ref{rem:scar} also apply.

\begin{corollary}
For $a < w < d$,
$$
\pi^*(w) > \pi^0(w),
$$
and for $w > d$,
$$
\pi^*(w) = \pi^0(w).
$$
In particular, $\pi^*(d-) > \pi^*(d+)$.
\end{corollary}

\section{Conclusions}\label{sec5}

We determined the optimal investment strategy for an individual who seeks to minimize her expected lifetime poverty, with a penalty for ruin, for a given {\it poverty} function, a non-negative, non-decreasing function of running wealth.  Our work is related to that in optimal investment to maximize expected utility of consumption or terminal wealth; however, we believe that it would be easier for an individual to choose a poverty function as compared with a utility function.  For that reason, we argue that minimizing expected lifetime poverty, with a penalty for ruin, is more objective than maximizing expected utility.

For the specific cases considered in Section \ref{sec30}, we proved that the optimal investment strategy increases with the poverty level $l$ and decreases with the penalty for ruin $\Pu$.  We expect these properties to hold for more general poverty and consumption functions.  That is, if $l_1(\cdot) \le l_1(\cdot)$ or if $\Pu_1 \ge \Pu_2$, then we expect $\pi_1(\cdot) \le \pi_2(\cdot)$.

For the specific cases considered in Section \ref{sec30}, we proved that, when wealth is above the poverty level $d$, the individual optimally invests as if she were minimizing her probability of lifetime ruin, an example of myopic investment.\footnote{We expect this myopia to hold for more general poverty and consumption functions.  That is, if $l(w) \equiv 0$ for $d < w < \ws$, for some $d \in (a, \ws)$, then we expect the optimal investment strategy will equal the one for minimizing the probability of lifetime ruin under the same consumption function.}  In many other goal-seeking problems, we have observed such myopia.

Indeed, Bayraktar and Young \cite{Bayraktar2007} found the optimal investment strategy to minimize the probability of lifetime ruin under constant consumption and under a no-borrowing constraint on investment, that is, the individual was not allowed to invest more in the risky asset than her current wealth.  Under that constraint, when the constraint did {\it not} bind (specifically, at greater wealth levels), then the individual invested as if the constraint did not exist.

More recently, Bayraktar et al.\ \cite{BPY2015} and Bayraktar and Young \cite{BY2015} determined the optimal investment strategy to maximize the probability of reaching a bequest goal with and without life insurance, respectively.  In the wealth regions for which it is optimal {\it not} to buy life insurance (specifically, at lower wealth levels), then the individual invested as if life insurance were not available.

Finally, Angoshtari et al.\ \cite{ABY2015b} minimized the expected lifetime spent in drawdown, that is, the time that one's wealth spends below some multiple of maximum wealth.  They showed that, when in drawdown, the optimal investment strategy is identical to the strategy for minimizing expected lifetime occupation of the same interval of wealth.  Furthermore, they showed that when the individual was not in drawdown, then the optimal investment strategy is identical to the strategy for minimizing the probability of lifetime drawdown.

We conjecture that myopic investment concerning constraints and opportunities is the rule, rather than the exception, in goal-seeking problems.

\bibliographystyle{plain}
\bibliography{bib_Asaf}

\begin{thebibliography}{10}

\bibitem{ABY2015b}
Bahman Angoshtari, Erhan Bayraktar, and Virginia~R. Young.
\newblock Minimizing the expected lifetime spent in drawdown under proportional
  consumption.
\newblock Working Paper, University of Michigan, 2015.

\bibitem{ABY2015a}
Bahman Angoshtari, Erhan Bayraktar, and Virginia~R. Young.
\newblock Minimizing the probability of lifetime drawdown under constant
  consumption.
\newblock Working Paper, University of Michigan, 2015.

\bibitem{Bayraktar2007}
E.~Bayraktar and V.R. Young.
\newblock Minimizing the probability of lifetime ruin under borrowing
  constraints.
\newblock {\em Insurance: Mathematics and Economics}, 41(1):196--221, 2007.

\bibitem{BPY2015}
Erhan Bayraktar, S.~David Promislow, and Virginia~R. Young.
\newblock Purchasing life insurance to reach a bequest goal while consuming.
\newblock Working Paper, University of Michigan, 2015.

\bibitem{Bayraktar2007b}
Erhan Bayraktar and Virginia~R. Young.
\newblock Correspondence between lifetime minimum wealth and utility of
  consumption.
\newblock {\em Finance and Stochastics}, 11(2):213--236, 2007.

\bibitem{BY2010}
Erhan Bayraktar and Virginia~R. Young.
\newblock Optimal investment strategy to minimize occupation time.
\newblock {\em Annals of Operations Research}, 176(1):389--408, 2010.

\bibitem{BY2015}
Erhan Bayraktar and Virginia~R. Young.
\newblock Optimally investing to reach a bequest goal.
\newblock Working Paper, University of Michigan, 2015.

\bibitem{BZ2015}
Erhan Bayraktar and Yuchong Zhang.
\newblock Minimizing the probability of lifetime ruin under ambiguity aversion.
\newblock {\em SIAM Journal on Control and Optimization}, 53(1):58--90, 2015.

\bibitem{Chakravarty2009}
Satya~R. Chakravarty.
\newblock {\em Inequality, Polarization and Poverty: Advances in Distributional
  Analysis}.
\newblock Springer-Verlag, New York, 2009.

\bibitem{Lambert2001}
Peter~J. Lambert.
\newblock {\em The Distribution and Redistribution of Income}.
\newblock Manchester University Press, Manchester, third edition, 2001.

\bibitem{Merton1992}
Robert~C. Merton.
\newblock {\em Continuous-Time Finance}.
\newblock Blackwell Publishing, Malden, Massachusetts, revised edition edition,
  1992.

\bibitem{MullShafir2013}
Sendhil Mullainathan and Eldar Shafir.
\newblock {\em Scarcity: Why Having Too Little Means So Much}.
\newblock Times Books, Henry Holt and Company: New York, 2013.

\bibitem{Royden1968}
Halsey~L. Royden.
\newblock {\em Real Analysis}.
\newblock Macmillan, New York, 1968.

\bibitem{Young2004}
Virginia~R. Young.
\newblock Optimal investment strategy to minimize the probability of lifetime
  ruin.
\newblock {\em North American Actuarial Journal}, 8(4):106--126, 2004.

\end{thebibliography}

\end{document}